\documentclass[aps,pre,twocolumn,superscriptaddress,showpacs]{revtex4-1}
\usepackage{graphicx}
\usepackage{amsmath}
\usepackage{natbib}
\usepackage[usenames]{color}
\begin{document}
\title{Symbolic walk in regular networks}

\author{L. Ermann}
\affiliation{Departamento de F\'{\i}sica, Gerencia de Investigaci\'on y Aplicaciones,
 Comisi\'on Nacional de Energ\'{\i}a At\'omica.
 Av.~del Libertador 8250, 1429 Buenos Aires, Argentina}

\author{G. G. Carlo}
\affiliation{Departamento de F\'{\i}sica, Gerencia de Investigaci\'on y Aplicaciones,
 Comisi\'on Nacional de Energ\'{\i}a At\'omica.
 Av.~del Libertador 8250, 1429 Buenos Aires, Argentina}

\date{\today}
\begin{abstract}
We find that a symbolic walk -- performed by a walker with memory given by a Bernoulli shift -- 
is able to distinguish between the random or chaotic topology of a given network. 
We show this result by means of studying the undirected baker network, 
which is defined by following the Ulam approach for the baker transformation in order 
to introduce the effect of deterministic chaos into its structure. The chaotic topology is 
revealed through the central role played by the nodes associated with the positions 
corresponding to the shortest periodic orbits of the generating map. They are the overwhelmingly 
most visited nodes in the limit cycles at which the symbolic walker asymptotically arrives. 
Our findings contribute to link deterministic chaotic dynamics with the properties of networks 
constructed using the Ulam approach.
\end{abstract}
\maketitle

\section{Introduction}
\label{sec:1}

The interest in the study of complex networks has been steadily increasing during recent years. 
Widespread use of the Internet and communication networks has undoubtedly been one of the major drivers behind this, 
however the wealth of applications they have are difficult to underestimate. In fact, networks research is interdisciplinary 
in nature, involving many areas \cite{NetworksBook1,NetworksBook2}. Their properties are relevant in sociology, epidemiology, 
economy and also physics, in particular, they could be useful for understanding transport phenomena in 
complex systems \cite{Transport1, Transport2}. Networks are usually classified in terms 
of the shape of the distributions of the node degree $k$, i.e., the number of connections associated 
to their nodes. Roughly speaking, one has regular networks where $k$ is a constant, also 
the ${\rm ER}$ model \cite{Networks1} where the distribution of $k$ is Poissonian, 
and finally scale free networks that have a power law distribution for their $k$ \cite{Networks2}.

On the other hand, characterizing the properties of transport has been an active field of research, 
with the study of random walks being of particular interest \cite{RWalks}. It is the common practice to consider 
a walker without memory that in fact leads to different kinds of diffusive processes, also 
depending on the nature of the medium. But in this work we are 
interested in studying one of a different kind, which we call {\em symbolic walk} (SW). One of the main motivations 
behind its definition is the possibility to open a two-way research route. On one hand they could be of help to devise 
more efficient protocols for message delivery in communication networks \cite{Communication,Communication2}. 
On the other, we could profit from networks properties in order to study transport in 
complex systems. The SW is performed by a walker with minimum information given 
by a short symbolic sequence derived from a Bernoulli shift \cite {Bernoulli}. 
In our case we consider 
a paradigmatic chaotic system, the baker map \cite{Baker,Baker2}. In this sense, the symbol sequence that gives 
the instructions on how to move to the walker can be associated to a periodic orbit of this map. 

We have taken the first step in determining the properties of this kind of walker by means of studying 
its behavior when moving in different types of regular networks. In doing so we can ask ourselves 
if this minimally informed walker is able to distinguish between the random or chaotic 
topology. For that purpose we have defined the baker (chaotic) network following the Ulam approach 
\cite{dimaUlam} of partitioning the phase space of the baker map in cells that can be associated to the nodes. This has 
the advantage of incorporating the chaotic features of the baker transformation into the network structure 
in a very simple way. In this paper we focus on regular networks since their simplicity allows us to 
study very interesting mathematical properties without losing the generality of the model we construct. 
We compare the SW behavior for the baker, random, and also for the ring and the torus networks. 

We have found that the SWs universally fall to what we call a limit cycle, i.e. a stationary 
path on the network that is repeated continuously, resembling the periodic orbits of dynamical 
systems. Moreover, we have verified that these limit cycles visit 
the nodes of the baker network associated with the shortest 
periodic orbits of the baker map with an overwhelming preference. This is not the case for the 
random network, where no such regularity is observed. 
We think that with the aid of this minimum information walker we could be able to design message delivery 
protocols that would profit from the chaotic topology of networks without the need of knowing their 
details. On the other hand, by studying global properties of complex networks we could be able to 
unveil features of transport in complex systems. Finally, we consider that these results can be 
expanded to different kinds of networks (not just regular) and also to the quantum realm.

The paper is organized as follows, in Sec. \ref{sec:2} we introduce the details of the symbolic 
walk and the baker network, we study its properties and compare with random regular, ring and torus networks, 
showing the main results of our work. In Sec. \ref{sec:3} we give an interpretation of these findings in terms 
of the shortest paths distributions and the connection among limit cycles on the network and the shortest periodic orbits 
of the baker map. Finally, in Sec. \ref{sec:4} we present our conclusions and ideas for future investigations.

\section{The SW in regular networks}
\label{sec:2}

\subsection{The SW}

A network can be explored in several ways. In a random walk each step of the walker at a given node is chosen at random from 
all the possible outgoing links. On the other hand, there are very complex paths, far from being random, that depend 
on multiple variables, as for example the ones followed by someone surfing the WWW or driving a car in a city.
Here, we will define the SW as a rule that the walker follows which is deterministic (having memory effects), trying 
to capture the essential features of this more realistic behavior.   

A network path can be characterized by the sequence of nodes that the walker passes through.
An alternative way to describe the path is given by the initial node, and the sequence of links followed.
We will take the latter approach in order to define the SW. For simplicity we will study only undirected 
and $k$ regular networks, i.e. networks where the number of neighbors per node is the same for all nodes.
This model could be extended to general networks in different ways, but this is out of the scope of the present work.

One of the most extended representations of networks is given by the $N\times N$ adjacency matrix 
\begin{equation}
A_{i,j}=\left\{ 
\begin{array}{cl} 
1& \text{link from $j$  to $i$} \\ 0& \text{otherwise}
\end{array} 
\right.
\end{equation}
In our case the number of non-zero elements per column and row is fixed to $k$.
An alternative representation for networks is given by a $N\times k$ link matrix
\begin{equation}
B(i,m)=j
\end{equation}
with $i,j=1,\ldots,N$ and $m=1,\ldots,k$ labeling the links between nodes $i$ and $j$. 
Note that the derivation of $B$ from the adjacency matrix $A$ is not unique, depending on link labels.
For simplicity we will write $B$ following the increasing order of the nodes.
Note also that undirected networks give rise to symmetrical $A$ matrices ($A_{i,j}=A_{j,i}$), while in $B$
this symmetry is expressed as the following: 
for each $m,i,j$ exists $m^\prime$ such that $B(i,m)=j$ and $B(j,m^\prime)=i$, 
where $i,j=1,\ldots,N$ and $m,m^\prime=1,\ldots,k$.

Following the $B$ representation, a path of $t$ steps can be characterized by an initial node and a strip of $t$ symbols, 
instead of the sequence of $t$ nodes  $(i_0i_1i_2\ldots i_{t-1}i_t)$. 
Using this notation, and given an initial node, we define the SW for a 
strip $\nu=\nu_1,\nu_2\ldots\nu_L$ of $L$ symbols (with $\nu_i=1,\ldots,k$)
as the deterministic evolution for infinite times with the sequence of neighbors given by the repetition of $\nu$.
Thus, one iteration of the SW with the strip $\nu$ is defined as:
\begin{equation}\label{def:sw}
 i_{t+1}=B(i_t,\nu_{t\ \text{mod}(L)})
\end{equation}

\begin{figure}
\includegraphics[width=0.475\textwidth]{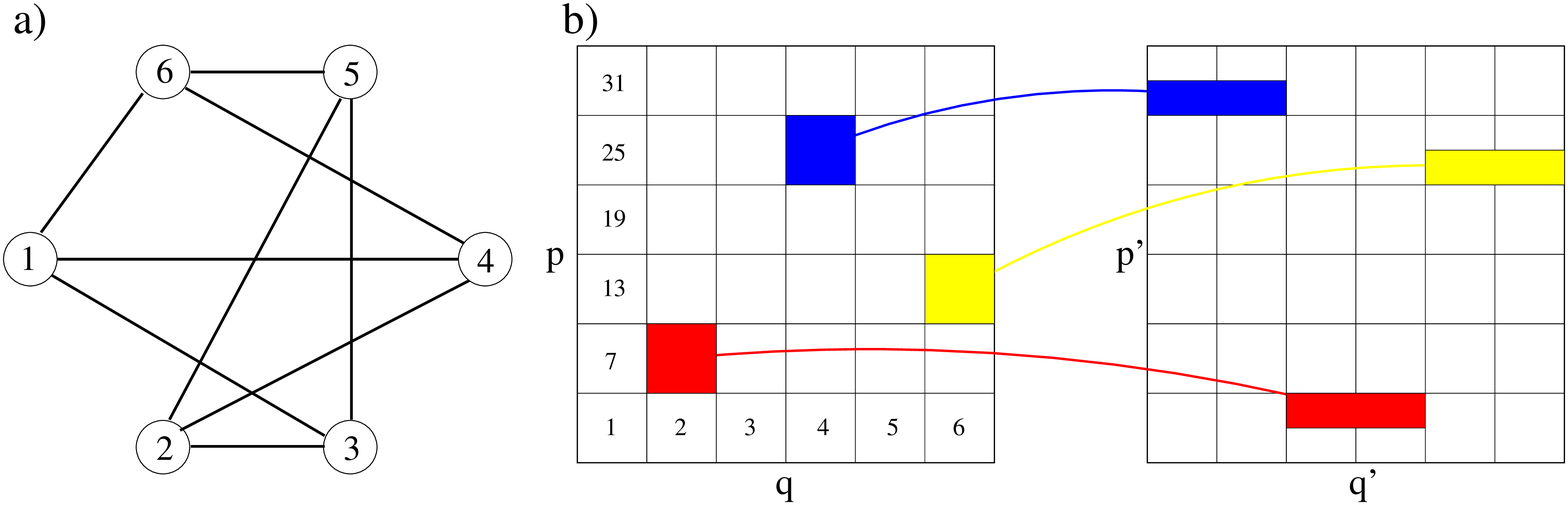}
\caption{(color online) The left panel ($a$) shows an example of a regular network of order $k=3$ with $N=6$.
The right panel ($b$) illustrates the \emph{baker network} defined by the Ulam procedure.
The phase space is divided in $N=N_q\times N_p$ cells ($N_q=N_p=6$ in this example) 
and the link between nodes are taken following the baker transformation. 
The label of each node is given by $i=i_q+(i_p-1)N_q$ where $i_q=1,\ldots,N_q$ and $i_p=1,\ldots,N_p$ 
Red, blue, and yellow initial areas (left panel) are evolved by the baker transformation 
(right panel) and define the links between nodes. 
The red area example shows the links between nodes $8-3$ and $8-4$.}
\label{fig:1}
\end{figure}

In order to clarify this definition we illustrate the SW with the following example: 
Fig. \ref{fig:1} a) shows a regular network with $N=6$ and $k=3$, which can be represented by means of the adjacency and 
link matrices $A$ and $B$ as
\begin{equation}
A=\left(
\begin{array}{cccccc}
        0&0&1&1&0&1\\
        0&0&1&1&1&0\\
        1&1&0&0&1&0\\
        1&1&0&0&0&1\\
        0&1&1&0&0&1\\
        1&0&0&1&1&0
\end{array}
\right); \ \ \ \ \ \ 
B=\left(
\begin{array}{ccc}
3&4&6\\
3&4&5\\
1&2&5\\
1&2&6\\
2&3&6\\
1&4&5
\end{array}
\right)\rm{.}
\end{equation}

The SW chosen for this example is the strip $\nu=12$.
Taking the initial node $i_0=2$, and following the SW definition (Eq. \ref{def:sw}), 
the walker takes the first node ($\nu_1=1$) and goes to the node $i_1=B(2,1)=3$.
The next step is $i_2=2$, and therefore the path is given by 
$2323232323\ldots$ in node representation. For this case the path could be seen as a periodic orbit of period $2$.
Choosing the initial condition $i_0=6$ we obtain the path $61414141414\ldots$, 
where the walker converges to what we call a {\em limit cycle} of period $2$ (again), but after some transitory evolution.
We will define two different quantities, $d_{t}$ and $d_{lc}$ which correspond to the transitory and the limit cycle 
dimensions respectively. For the dimension of the transitory we take into account all nodes belonging to the path of the 
walker from the initial node. Note that $d_{t}\ge d_{lc}$, since the limit cycle is included in the transitory definition, 
and then we have $d_{t}=d_{lc}=2$ for the first initial condition, and $d_{t}=3, d_{lc}=2$ for the latter case.

In this example we only have two different limit cycles with $d_{lc}=2$ given by $lc=14$ 
(or equivalently $lc=41$) for initial conditions $i_0=3,4,5,6$,
and $lc=23$ for initial conditions $i_0=1,2$. The transitory dimensions in all cases 
are $d_t=2$ for $i_0=2,4$, $d_t=3$ for $i_0=1,3,6$, and $d_t=4$ for $i_0=5$.

\subsection{Regular networks}

Since the SW is expected to behave differently depending on the kind of network it moves in, we 
consider its dynamics in chaotic, random and also in simple networks. In the following we describe 
their construction.

The definition of a \emph{chaotic} network follows the Ulam method for chaotic maps \cite{dimaUlam,dimaUlam2,dimaUlam3,dimaUlam4}. 
For chaotic networks we have chosen the baker map that has a uniform Lyapunov exponent in the unit square phase space.
The Ulam approach for the baker transformation consists of dividing the phase space in  $N=N_q\times N_p$ cells, each one representing 
a node whose label is given by $i=i_q+(i_p-1)N_q$. A link between nodes is added if the baker transformation 
(or its inverse since our network is an undirected one) maps the node $i$ to the node $j$ as illustrated in Fig\ref{fig:1} b).
The usual baker transformation is defined as
\begin{equation}
\begin{array}{ccc}
q^\prime&=&2q-[2q]\\
p^\prime&=&p/2+[2q]
\end{array}
\end{equation}
where $[x]$ is the integer part of $x$.
The degree of the baker network is  $k=2\exp{\lambda}$  with $\lambda=\ln 2$, and therefore $k=4$ for this baker map. 
Note that this 2-baker transformation can be easily generalized to  an $n$-baker one 
(i.e. with $n$ partitions of the phase space) in order to obtain networks with different orders.

The {\em random} network construction is done in the usual way, i.e. establishing a two-way link between a random pair of nodes $i$ and $j$ 
and following in that sense until each node has $k$ links. Finally, we define two different simple regular networks given by the 
\emph{torus} and \emph{ring} topology. The \emph{torus} network with $k=4$ is defined by means of a two dimensional torus of 
$N=N_q\times N_p$ nodes with $k=4$ neighbors each, given by left, right, up and down links. On the other hand, 
in the \emph{ring} network all nodes are in a ring, having $k$ links with their closest neighbors. For the case of $k=4$ 
each node has links with their first and second closest nodes per side. Note that both simple models can be generalized 
to other values of $k$.

\begin{figure}
\includegraphics[width=0.475\textwidth]{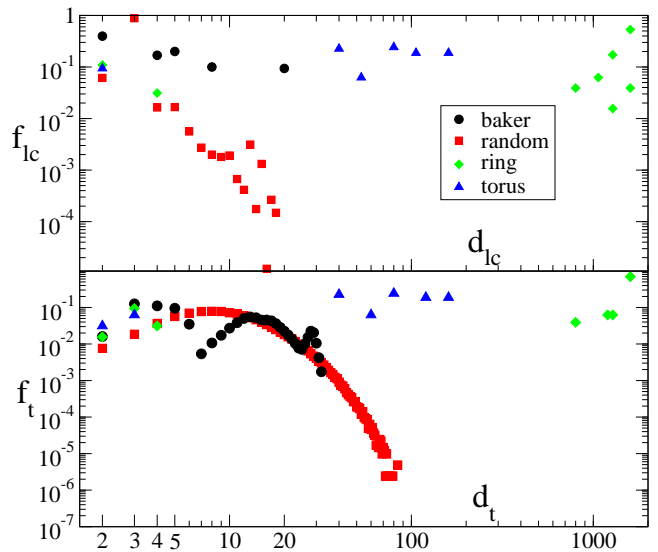}
\caption{(color online) Fraction of paths with dimension $d_{lc}$ and $d_t$ ($f_{lc}$ and $f_t$),
on top and bottom panel respectively. The number of nodes is $N=6400$, and the statistics is carried out 
for all strips $\nu$ of dimension $L=4$ and all initial conditions ($i_0=1,2,\ldots,N$).
The regular networks used (with $k=4$) are the baker, random (5 realizations), ring and torus represented by 
black circles, red squares, green diamonds and blue triangles, respectively.}
\label{fig:2}
\end{figure}

Now that we have defined the main tools used in our investigation we can show the results. 
The first step is to analyze the statistical properties of the transitory and limit cycles for the SW dynamics.
We have fixed the degree of the regular networks at $k=4$ in order to have a simple but non-trivial evolution.
We start the evolution from all initial nodes $i_0=1,\ldots,N$, using all symbolic strips $\nu$ of dimension $L$.
It is worth noting that when $L$ is a multiple of $k$, the statistics obtained by covering all these possibilities is 
independent on the order of the nodes and links. For this reason we take $L=k=4$, which include the strips of periods 
$L=1$ and $L=2$. Therefore, the number of paths taken into account for the full statistics is $N_t=N*k^L=N*4^4=256N$. 
The fraction of paths that have transitory and limit cycle dimensions $d_{lc}$ and $d_t$  
are defined as $f_{lc}=\#_{d_{lc}}/N_t$ and $f_t=\#_{d_{t}}/N_t$, respectively.
The results for $N=6400$ are shown in Fig. \ref{fig:2} for the baker, random, ring and torus networks.
The first thing we notice is that the limit cycles of the baker network have just $5$ different dimensions in clear 
contrast with a whole spectrum for the random network, which on the other hand stretches in more or less the same 
range. If we look at the behavior of the simple networks they resemble the one of the baker network in having 
limit cycles with a few dimensions, but the range of them is quite different, being larger for the simple network 
with larger diameters, i.e. the ring. It seems that the chaotic network shares properties from both simple and 
random ones. When looking at the transitory behavior the simple networks show no new information which corresponds 
to the fact that their simple topology induces a very regular motion that makes almost coincide both regimes. 
In the case of the baker network the dimensions of tansitories now cover a wide spectrum while staying in the range 
of the limit cycles, contrasting once again with the random network, whose dimension spectrum behaves smoother 
and extends to values much larger than in the former case. 

\begin{figure}
\includegraphics[width=0.475\textwidth]{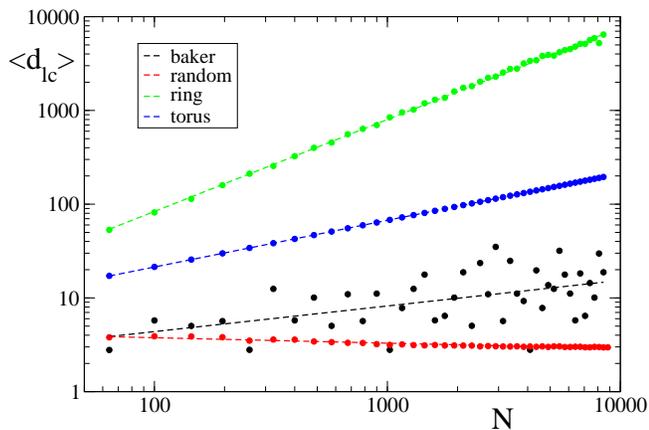}
\caption{(color online) Average limit cycle dimension $\langle d_{lc}\rangle$ as a function of network size $N$ for different networks.
Results for the baker, random, ring and torus networks are represented by black, red, green and blue circles, respectively. 
The average is computed over all SWs (full statistics) with $L=4$, and over 5 random realizations in the random network case.
Dashed lines correspond to the power law $\langle d_{lc}\rangle \propto N^{\alpha}$, with the fitted exponents: $0.98\pm0.03$ for the ring, 
$0.50\pm0.03$ for the torus, $0.27\pm0.09$ for the baker, and $-0.06\pm0.04$ for the random networks.}
 \label{fig:3}
\end{figure}

\begin{figure}
\includegraphics[width=0.475\textwidth]{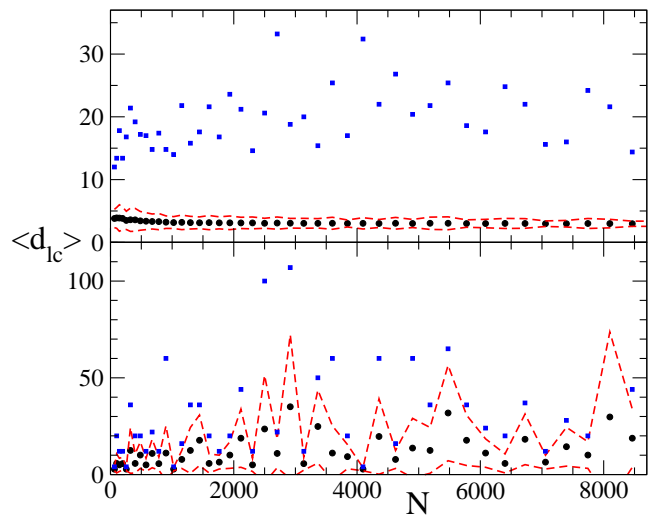}
\caption{(color online) Black circles represent the average limit cycle dimension $\langle d_{lc}\rangle$ as a function of 
network size $N$ for the random and baker networks in top and bottom panels, respectively. The calculation corresponds to 
all SWs (full statistics) with $L=4$, and over 5 random realizations in the random network case. The blue squares represent 
the maximum limit cycle dimensions, while dashed red lines show the standard deviation $\langle d_{lc}\rangle\pm\sigma$.}
\label{fig:4}
\end{figure}

In order to better understand this we have calculated the way in which the limit cycles dimensions behave 
with respect to the size of the network; results can be seen in Fig. \ref{fig:3}. 
The average limit cycle dimension approximately follows a power law behavior with respect to the network size $N$ in all cases, 
i.e. $\langle d_{lc}\rangle \propto N^{\alpha}$. The greatest exponent belongs to the ring network for which we have fitted 
a value $\alpha=0.98\pm0.03$, then comes the torus network with $\alpha=0.50\pm0.03$, which can be explained by the fact 
of having a smaller diameter. For the baker network $\alpha=0.27\pm0.09$, but the fluctuations are larger than 
in all the other cases, confirming how relevant the chaotic structure of the map phase space could be for the corresponding network topology. 
Finally, the random network behavior is characterized by $\alpha=-0.06\pm0.04$, which indicates that the average of limit cycle 
dimensions is almost independent on network size.
This different behavior clearly distinguishes it from the chaotic network, and can be associated to the local properties 
of the random construction. A deeper analysis of the $\langle d_{lc}\rangle$ growth for the random and baker network cases can be seen 
in Fig. \ref{fig:4}. The values of $\langle d_{lc}\rangle$  for the random network (top panel) converges to 
$\langle d_{lc}\rangle\sim3$. This can be explained due to the typical orbit of 3 nodes given by $ii^\prime ii^{\prime\prime}$
which dominates the statistics. 
It is even more evident now, how the SWs are able to distinguish between both kinds of networks, being their average limit cycle 
dimensions behavior strikingly different. While regularity is the norm in the random case, strong fluctuations are the main 
feature in the chaotic one, not only shown by the standard deviation $\langle d_{lc}\rangle\pm\sigma$ values, but also 
by the maximum values represented by means of blue squares. With the aid of Fig. \ref{fig:4} we are able to underline the 
marked dependence of the baker network structure on the properties of the chaotic map phase space, whose coarse-graining effects given 
by the size $N$ can induce abrupt changes in the links structure.
Another interesting conclusion extracted from Fig. \ref{fig:4} is that the maximum $d_{lc}$ is 2 to 5 times larger for the baker network 
when compared to the random case. 

\begin{figure}
\includegraphics[width=0.475\textwidth]{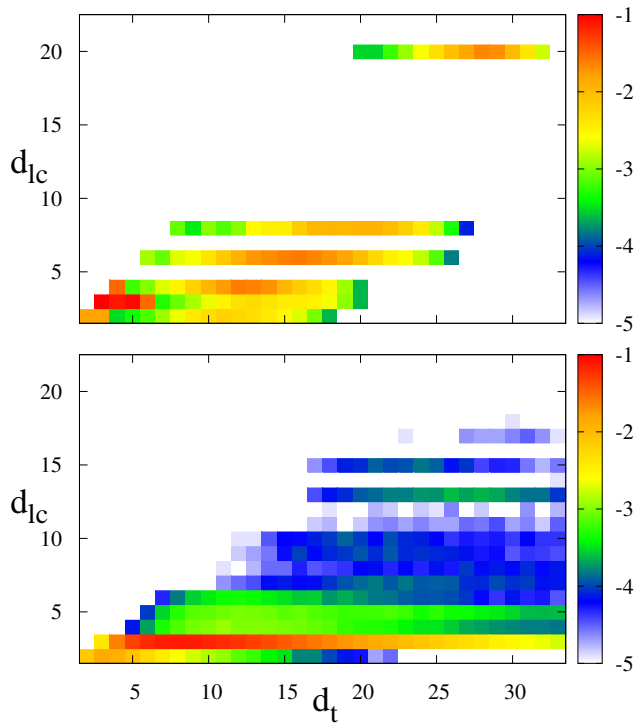}
\caption{(color online) Joint probability $f(d_t.d_{lc})$ for a path to have a limit cycle and transitory 
dimensions $d_{lc}$ and $d_t$, in logarithmic scale. Top and bottom panels show the cases of the baker 
and one realization of the random networks, respectively with $N=6400$ and full statistics with $L=4$.}
\label{fig:5}
\end{figure}

Finally, we present a very nice way to see the ability of the SWs to tell the difference between the baker and random networks. 
The joint probability (or fraction) for a path to have both $d_{lc}$ and $d_t$ 
dimensions $f(d_t.d_{lc})$ is shown in Fig. \ref{fig:5} in logarithmic scale for the baker network in the top panel and 
for the random network in the bottom one.
In this figure it is visible how the limit cycle of dimension 3 dominates the statistics for both cases, 
but it is even stronger in the random case. Typical transitory dimensions are larger for the random case,
and there is a correlation between large values of $d_{lc}$ and $d_t$ for the baker network.


\section{Limit cycles and short periodic orbits}
\label{sec:3}

In Sec. \ref{sec:2} it has become clear that SWs are able to distinguish between the different random or chaotic 
topology of a given network. But, why is this so? In this Section we are going to try to answer this 
question. 

\begin{figure}
\includegraphics[width=0.475\textwidth]{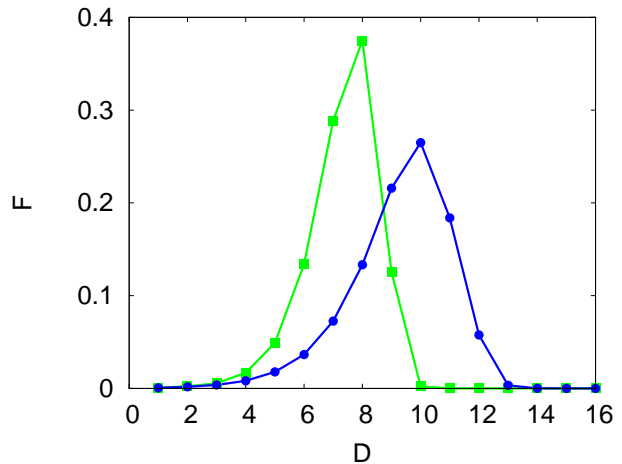}
\caption{(color online) Histograms showing the fraction $F$ of the shortest source-target distances $D$ 
(measured in number of nodes) for the baker and random networks. The (green) gray line with squares 
corresponds to the baker network while the (blue) black line with circles corresponds to the random 
regular network. Both cases were calculated for $N=6400$, taking into account all possible 
source-target pairs for the baker network, and $10$ realizations and a $0.1$ fraction of all possible 
pairs (taken at random) for the random network.}
\label{fig:6}
\end{figure}

As a first step we have measured the source-target distances of both the baker and the random networks in order 
to see if there is a noticeable difference in the way a given pair of nodes are connected. We have used the 
standard breadth-first search algorithm in order to find the shortest distance for all pairs of nodes 
in the baker network, and $10$ realizations with a $0.1$ fraction of all possible pairs for the random 
network. In Fig. \ref{fig:6} we show histograms considering $N=6400$.
There is a significant shortening of the source-target distances for the chaotic network (represented by means 
of (green) gray line with squares) when compared to the random one ((blue) black line with circles). 
This could be the reason behind the longer transitory regimes observed for the random case but it is not enough 
to explain the so limited number of dimensions for the limit cycles found in the baker network.

\begin{figure}
\includegraphics[width=0.475\textwidth]{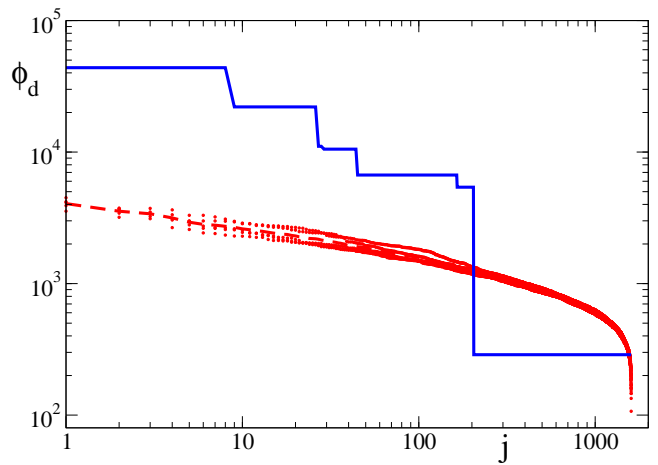}
\caption{(color online) Number of limit cycles that pass through the node $j$: $\phi_{\rm d}(j)$ as a function of the 
(reordered) node label $j$, for the baker and the random networks with $N=1600$ and $L=4$. Results for the baker network are 
represented by means of blue lines, while those for the random network (calculated for 5 realizations) by means of 
red circles (each instance) and a dashed line (average), respectively.}
\label{fig:7}
\end{figure}

In order to find out why there are so few different dimensions we begin by showing the number of limit cycles  
that pass through the node $i$: $\phi_{\rm d}(i)$. This quantity is shown in decreasing order 
($\phi_{\rm d}(j)\ge\phi_{\rm d}(j+1)$) in Fig. \ref{fig:7}.
Remarkably, there are huge degeneracies for this quantity in the baker network case while not in the random one (the 
behavior for this latter being quite smooth). 

\begin{figure}
\includegraphics[width=0.475\textwidth]{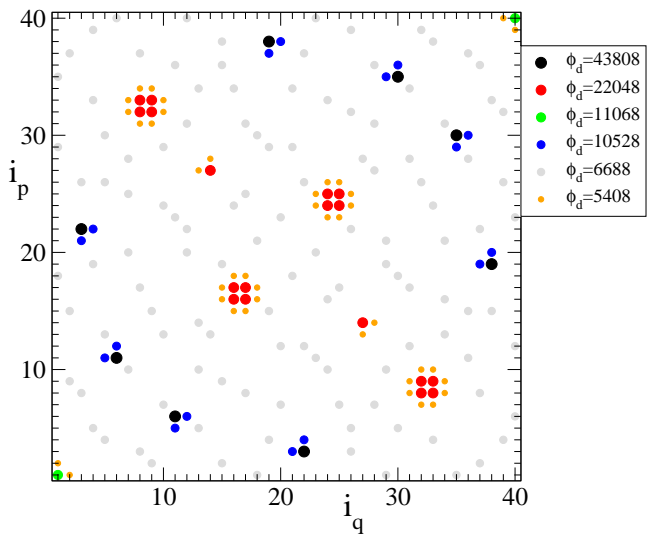}
\caption{(color online) Phase space representation of the nodes with largest $\phi_{\rm d}$ for the baker network 
with $N=1600$ ($N_q=N_p=40$) and $L=4$. Different symbols in the phase space lattice correspond to 
the following different values of $\phi_{\rm d}=43808,22808,11068,10528,6688,5408$. For nodes without symbols $\phi_{\rm d}=288$. 
The largest values of $\phi_{\rm d}$ can be associated with the shortest periodic orbits of the baker map.}
\label{fig:8}
\end{figure}

But, can we ascribe these degeneracies to some structure behind the nodes highly preferred by limit cycles? It turns out that we can. 
These nodes are associated with the shortest periodic orbits of the baker map that we have used to generate the baker 
network following the Ulam procedure (see Fig. 2 of \cite{Baker2} for a phase space representation of shortest periodic orbits). 
This is nicely shown in Fig. \ref{fig:8} where we represent the nodes that have the largest values of $\phi_{\rm d}$ in the phase space 
of the baker map by means of circles with different size and color. 
The nodes given by green and orange circles can be associated to the fixed points, those by red and orange circles to the symmetric 
periodic orbit of period 2 and 4, represented in symbolic dynamics as $01$ and $0011$ respectively, while black and blue 
circles are related to the periodic orbits of period 4 $0001$ and $0111$.


\section{Conclusions}
\label{sec:4}

In this work we have introduced a new kind of walker which is a deterministic alternative to the random one. We have 
called the associated dynamics a SW, since the steps of the walker are dictated by a symbolic sequence 
coming from a Bernoulli shift that can be associated to the periodic orbits of a chaotic system. In our case we have 
considered the paradigmatic baker map. 

We have found that this minimal information walker can distinguish between a chaotic network and a 
random one. This is shown by means of defining the baker (chaotic) network through the Ulam procedure which identifies 
a node with a given area in phase space. The SW turns out to always 
fall into what we call limit cycles, which in this context are closed paths. Furthermore, in the baker network case, these closed paths show an 
overwhelming preference to visit the nodes associated with the shortest periodic orbits of the baker map. This is in clear 
contrast with what happens for random regular networks, where no such asymptotic behavior has been found.  

In this sense, SWs show themselves as a promising idea in the seek of efficient protocols for message 
delivery in communications networks. As a matter of fact it seems possible to engineer a simple series of instructions 
taking into account the global properties of a network in order to have a given path as the target. Moreover, 
this strong link between the properties of Ulam networks and periodic orbits of chaotic systems could lead to a 
better use of the knowledge coming from both areas, in particular for investigating transport phenomena in complex systems.

In the future \cite{future} we will extend this calculations to more kinds of networks (other than regular and based on other systems). 
We will also explore the way in which one can select different limit cycles by designing specific protocols. Furthermore, we 
will study the transport properties of SWs beginning with their first passage times behavior \cite{mfpt}. 
Finally, we plan to extend this calculations to quantum graphs and maps .

\section{Acknowledgments}
Financial support form CONICET is gratefully acknowledged. 
We thank Marcos Saraceno for useful discussions.

%

%

\begin{thebibliography}
\eprint{}

\bibitem{NetworksBook1}
M. Newman, 
\textit{Networks: An Introduction}, 
(Oxford University Press, New York, 2010).

\bibitem{NetworksBook2}
R. Cohen and S. Havlin, 
\textit{Complex Networks: Structure, Robustness and Function}, 
(Cambridge University Press, Cambridge, 2010).

\bibitem{Transport1}
L. Skarpalezos, A. Kittas, P. Argyrakis, R. Cohen, and S. Havlin, 
Phys. Rev. E \textbf{88}, 012817 (2013).

\bibitem{Transport2}
R. Guimer\'a, A. D\'\i az-Guilera, F. Vega-Redondo, A. Cabrales, and A. Arenas, 
Phys. Rev. Lett. \textbf{89}, 248701 (2002).

\bibitem{Networks1}
P. Erd\"os and A. R\'enyi, 
Publ. Math. \textbf{6}, 290 (1959).

\bibitem{Networks2}
R. Albert, H. Jeong, and A.-L. Barab\'asi, 
Nature (London) \textbf{401}, 130 (1999).

\bibitem{RWalks}
G.H. Weiss, 
\textit{Aspects and Applications of the Random Walk}
(North-Holland, Amsterdam, 1994).

\bibitem{Communication}
R. Beraldi, R. Baldoni, and R. Prakash, 
IEEE Transactions on Mobile Computing \textbf{9}, 1649 (2010).

\bibitem{Communication2}
J.P. C\'ardenas, M.L. Mouronte, L.G. Moyano, M.L. Vargas, and R.M. Benito, 
Physica A \textbf{389}, 4209 (2010).

\bibitem{Bernoulli}
P. Gaspard, 
\textit{Chaos, Scattering and Statistical Mechanics}
(Cambridge University Press, Cambridge, 2005).

\bibitem{Baker}
M. Saraceno, 
Ann. Phys. (Leipzig) \textbf{199}, 37 (1990).

\bibitem{Baker2}
L. Ermann and M. Saraceno, 
Scholarpedia \textbf{7} (12), 9860 (2012).\\ 
\verb|www.scholarpedia.org/article/Quantized_baker_map|

\bibitem{dimaUlam}
L. Ermann and D.L. Shepelyansky, 
Phys. Rev. E \textbf{81}, 036221 (2010); 
S.M. Ulam, 
\textit{A Collection of mathematical problems, Vol. 8 of
Interscience tracs in pure and applied mathematics} 
(Interscience, New York, p. 73 ,1960).

\bibitem{dimaUlam2}D.L. Shepelyansky and O.V. Zhirov, 
Phys. Rev. E \textbf{81}, 036213 (2010). 

\bibitem{dimaUlam3}K.M. Frahm and D.L.Shepelyansky, 
Eur. Phys. J. B \textbf{76}, 57 (2010). 

\bibitem{dimaUlam4}L. Ermann and D.L. Shepelyansky, 
Eur. Phys. J. B \textbf{75}, 299 (2010).

\bibitem{future}
L. Ermann and G.G. Carlo, unpublished.

\bibitem{mfpt}
J.D. Noh and H. Rieger
Phys. Rev. Lett. \textbf{92}, 118701 (2004). 

\end{thebibliography}
\end{document}